# Localizing Audiences' Gaze using a Multi-touch Electronic Whiteboard with sPieMenu

Kazutaka Kurihara   Naoshi Nagano   Yuta Watanabe
Yuichi Fujimura   Akinori Minaduki   Hidehiko Hayashi   Yohei Tsuchiya
National Institute of Advanced Industrial Science and Technology (AIST)
Akihabara Daibiru 11F, 1-18-13 Sotokanda Chiyoda-ku Tokyo 101-0021 JAPAN
k-kurihara@aist.go.jp

**ABSTRACT**
Direct-touch presentation devices such as touch-sensitive electronic whiteboards have two serious problems. First, the presenter's hand movements tend to distract the audience's attention from content. Second, the presenter's manipulation tends to obscure content. In this paper we describe a new electronic whiteboard system that supports multi-touch gestures and employs a special pie menu interface named "sPieMenu." This pie menu is displayed under the presenter's palm and is thus invisible to the audience. A series of experiments shows that the proposed system allows both novice and expert users to efficiently manipulate the electronic whiteboard, and that the proposed system decreases distraction to the audience compared to traditional approaches.

**ACM Classification Keywords**
H.5.2 User Interfaces: Input devices and strategies.

**INTRODUCTION**
In recent years big-screen displays and direct-touch input devices, which are frequently used to describe and directly interact with presentation materials as an aid to explanation and debate, have become increasingly commonplace. They are now frequently encountered in classrooms, presentations, and meetings, and often involve audiences of unspecified size. These systems have great potential as they realize the direct communication advantages of the well-known traditional blackboard (whiteboard) and chalk approach, while making it possible to incorporate multimedia materials and utilize other information and communication technologies.

From the perspective of HCI there is an important interface design issue that must be considered when designing electronic whiteboard systems that combine big-screen displays and direct-touch input devices. This is the problem of localizing the function selection menu [10]. In the case of a traditional desktop application, if the function menu is located at the perimeter of the application window, then as the size of the screen grows so too does the cost associated with shifting the focus of the application (the cursor for example) to the menu. This issue is relevant not only to large-screen displays but also to tablet PCs and other devices that support direct-touch input and thus consume greater energy in order to shift or move the cursor.

Several other authors have addressed this issue, including Callahan et al. [1], Ramos et al. [10], Grassman et al. [11], who introduced alternative function selection interface modalities including the pie menu, pressure widget, and hover widget which rely on relative displacement from the current cursor position, pen pressure, and pen height respectively. These approaches may be said to achieve interface localization by supporting interactions that do not depend on the user's current absolute cursor position.

In this research, we extend the interface localization paradigm, and in particular, investigate the idea of "audience gaze localization." The works described above focus on a 1-to-1 relationship between the user and the application, with the aim to optimize the utility for the user given this situation. In the case of a classroom-oriented electronic whiteboard, a generic presentation tool, or a business meeting support tool however, in fact the interaction normally involves one user or presenter who is interacting through the application with an unspecified number of people in an audience, and it is necessary to carefully consider this fundamental difference when designing the interface. Here, in addition to maximizing the usability of the application with respect to the user, it is also necessary to consider an approach that minimizes the negative effects of the user's manipulations from the perspective of the audience. For example, if in manipulating the menu the user is frequently blocking the view of the presentation contents with his torso, arms or fingers, this may cause the audience to focus not on the information being presented but rather on the movements of the presenter. Alternatively if in order to select a new item or function, the user is required to open a large menu in the display, this may obscure the presentation contents, or inadvertently draw the audience's attention to the menu itself.

Localizing audience gaze is one approach to dealing with these issues, where user's manipulations and corresponding body movements, as well as changes in the information being displayed can inadvertently or unnecessarily distract the audience from the main communication goal, thus this work proposes an approach to interface design that seeks to minimize these negative effects.

In this work we report on the recent development of an electronic whiteboard system enabled with multi-touch technologies as a means to effectively realize audience gaze localization. The system employs the frequently used multi-touch gesture based approach and combines this with a specially designed pie menu that we refer to as "sPieMenu" in order to achieve menu localization. The sPieMenu differs from a traditional pie menu in that it is automatically hidden from the view of the audience under the palm of the user's hand. This makes it possible to support a wide variety of user-level functions that would be difficult to achieve through multi-touch gestures alone, while also serving as a design approach to achieving audience gaze localization.

The remainder of this work is structured as follows. Chapter 2 provides an outline of related background research, while Chapter 3 describes the development and implementation of the proposed multi-touch electronic whiteboard. Chapter 4 describes several experimental evaluations and provides additional discussion, and Chapter 5 concludes the paper.

## RESEARCH BACKGROUND

### Direct-Touch Electronic Whiteboards

When designing an interface for an electronic whiteboard system it is important to consider not only usability issues from the perspective of the user, but also the impact of the user's actions and their results on the audience. Electronic whiteboard systems driven by direct-touch interfaces have become increasingly popular in recent years, largely because these devices present advantages for both the presenter and the audience. From the perspective of the former, a direct-touch interface presents a much more intuitive, direct interaction in pointing and dragging operations compared to traditional modalities such as a mouse interface. From the perspective of the latter, materials that are presented via direct-touch help the audience better understand the focus of the presenter and thus follow along with the presentation more easily [2].

Nevertheless direct-touch electronic whiteboards also present a difficult problem. The interfaces for these whiteboards tend to have relied on toolbars and buttons that are usually located at the periphery of the application window. In informal experiments where we employed a gaze tracking system to track the gaze of a test audience, it was observed that actions that manipulated buttons located at the edge of the screen tended to obscure the presentation materials and also resulted in the audience's gaze being strongly affected by elements other than the presentation materials themselves (Figure 1). Moreover, the subjects in these experiments also commented that the presenter's arm movements were distracting.

An effective interface design approach for direct-touch electronic whiteboard systems should focus on minimizing or eliminating situations where the user's actions inadvertently obscure the presentation materials, distract the audience or otherwise draw the audience's attention to the actions themselves.

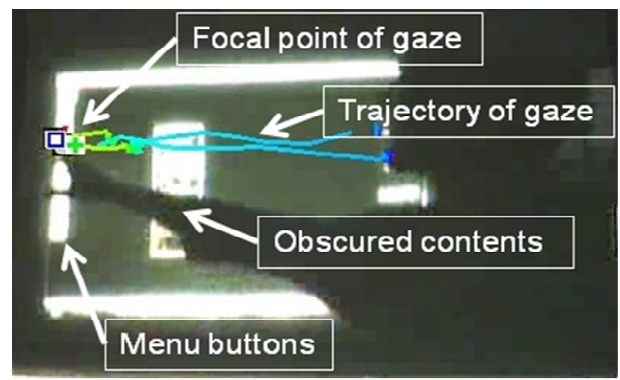

**Figure 1. Analysis of audience gaze movement for an electronic whiteboard**

### Multi-touch Gestures

A multi-touch gesture interface makes it possible to utilize rich gestures through the use of multiple fingers and thus to realize a wide variety of actions with a minimum of movement, all of which makes this a very attractive interface approach for an electronic whiteboard system.

Oguni et al. [7] discussed the possibility of single-touch gestures for an electronic whiteboard system, however they did not show an implementation or adopt the approach as they concluded that the gestures were non-intuitive and too difficult for users to remember. The main reason for this is that a single touch interface, such as that supported by a single finger or a pen, is only capable of recognizing a single point of movement. This means that the only way to support multiple functions through such an interface is to increase the length and/or complexity of individual gestures. As they become more complex these gestures increasingly diverge from people's natural movements, and thus become more difficult for user's to learn.

If a device that supports multi-touch gestures is employed, multiple fingers or both hands can be used. It makes possible gestures that employ multiple input points or movement in multiple simultaneous directions. This combination in turn makes it possible to support a wide variety of different functions using simple, natural movements. People's intuitive, "I want to do this," cognitive impulses can be linked to operational actions [6], making it easier for users to memorize the associated gestures. This approach also aids in producing a sense of successful interaction from the perspective of the user [6]. Looking at the problem in this light it is easy to see how natural gestures like opening or closing a hand, which are easy for people to learn and remember, can be linked to operations like moving the process state forward or backwards in a timeline, moving up, down, right, or left, and enlarging or shrinking as well as other physical concepts related to "direction," "time," or "size". On the other hand, operations like changing a pen color or printing, which do not lend themselves naturally to physical representations, can still be difficult to remember when encoded as gestures. Even when considering just those functions that are frequently used in connection with electronic whiteboards; stroke or brush based painting, changing the color or width of a pen, shrinking or enlarging photographs, directional movement, copy and paste operations, enlarging or shrinking a drawing canvas, changing slides, undo, redo, printing, etc., it would not necessarily be user friendly to encode all of these actions as gestures. In response to this challenge we investigate combining these multi-touch gestures with the menu selection interface described in the following section.

In regard to multi-touch gesture interfaces, much practical progress has been made through commercial devices such as the Microsoft Surface, Touch Diamond, and iPhone, etc. toward developing, standardizing and modeling common gestures [12]. In this paper we focus in particular on a new hybrid interface that integrates existing multi-touch gestures with a new menu-selection interface, and not on the multi-touch gestures themselves.

**Localization of the Menu Selection Interface**
The pie menu interface has previously been proposed as one method for implemented menu-selection localization with a direct-touch interface. In this approach, at the point where the pen or finger touches the screen, a pie-shaped menu is opened using the point of contact as its center. Unlike a traditional pull-down menu, where items are arranged vertically, a pie-shaped menu has the advantage that each of the menu items is equidistant from center point, which reduces the amount of movement required for the user [1] to perform a selection. Moreover, this can be cleverly and effectively combined with smooth transitions to gesture-based input (called *marking menu*) that does not require displaying the menu [13].

Recently, menu selection interface localization approaches have begun to take advantage of the redundant degrees of freedom afforded by recent improvements in hardware. The Pressure Widget [10] applies pressure sensing technology to a stylus pen, while the Hover Widget [11] employs technology capable of detecting the hover state of a similar pen.

Nevertheless, these menu-selection interfaces for electronic whiteboard systems still display the menu itself to the audience, and this has been shown to distract the audience. Thus it is important to consider the problem that this obfuscation represents as we do in the following section

**Handling Occlusion Problem**
One of the major issues related to direct-touch interfaces is the "occlusion problem". The user or presenter tends to hide or obscure the actual information being presented with their body, actions or through opening up a menu. In [9] Vogel et al. modeled the degree to which pen actions tend to obscure the display area in the case of a tablet PC. In [8] Brandt et al. developed an interface approach for a multi-touch display which estimates the position of the user's hand in relation to the touch-screen and dynamically places the menu so as to ensure that it is not obscured by the user's hand.

In this paper we focus on making creative use of the obstructed space. Electronic whiteboard systems are usually used to display presentation material to an audience. In this scenario the presenter (user) may be using their body to point out or explain the materials, however from the perspective of the audience this is an obstruction. This area, which represents an obstruction to the audience but which is visible to the user may be used to construct interaction methods so that presentation contents are not hidden from the audience while functional operations are secretly available.

**MULTI-TOUCH ELECTRONIC WHITEBOARD**
In this chapter we discuss the construction and design of the proposed multi-touch electronic whiteboard. The proposed system is based on general-purpose multi-touch capable hardware and electronic whiteboard software, as well as the interaction interface associated with it.

The multi-touch panel employs the Frustrated Total Internal Reflection (FTIR) system developed by Han [3] as its baseline hardware. For the electronic whiteboard software, the open source presentation platform "kotodama" [4] is employed.

**Interface Outline**
In the context of a direct-touch electronic whiteboard system, gesture-based operations will be an effective means to reduce arm movement for the user while focusing the attention of the audience on the contents of the presentation. Also, because some gestures are difficult to learn and remember it is important to consider assistive functions to help reduce cognitive load. Nevertheless it is also important to ensure that these assistive functions do not unnecessarily distract the audience or obscure the presentation materials.

In order to achieve these goals, below we propose a new electronic whiteboard interface that combines gesture-based controls with the new sPieMenu.

*Gesture-based Controls*

In the context of actual presentations involving electronic whiteboards, the most frequent actions that need to be considered are: stroke-based input, moving objects, and enlarging or shrinking objects. As above described (in "Multi-touch Gesture" section), these actions were associated with one or two finger gestures according to human behavioral cognition information. Other functions were associated with either three fingered gestures or the sPieMenu interface.

- 1-fingered movement: stroke-based input (Figure 2)
- 2-fingered parallel movement: select and move an existing object, or touch the background area to select and move the entire canvas (Fig. 3)
- 2-fingered open-close: enlarging or shrinking an object, or enlarging or shrinking the entire canvas area over a background (Figure 4)
- 3-fingered rotation: undo, redo (Figure 5)

**sPieMenu**

The sPieMenu displays the pie menu underneath the user's palm thus allowing the user to easily check the available actions (Figure 6). By using the multi-touch panel it is possible to display the sPieMenu only when the user touches the panel with the specified number of fingers. The main purpose of the sPieMenu then is to allow the user to check available actions, while avoiding unnecessary distraction or obstruction of the presentation materials from the perspective of the audience. Furthermore, as the user increases their level of expertise actively displaying the menu becomes unnecessary and the system can be shifted to a simple multi-touch gesture approach. In contrast to the transition from a typical pie menu to a marking menu, in this case no explicit transition from the display state to the non-display state is actually necessary. This aspect could be particularly useful for "in class student presentations" where numerous users of varying skill share the same electronic whiteboard system.

The specific actions are described below. If the user touches the display with three fingers, the system automatically computes the orientation and location of the hand from the three points of contact and displays the pie menu under the user's palm. As long as the fingers are touching the panel, the menu will be displayed at a fixed distance from the user's fingers. Thus even if the user moves his arm, the pie menu will always be hidden beneath his palm, and thus hidden from the view of the audience (Figure 6). The parameters that govern the distance from the contact point where the menu should be centered, as well as the menu orientation, may be customized by the user if necessary.

Following the three-fingered touch, and depending on the direction of the subsequent movement, the following actions are possible:

- Left: Go to the previous slide (Back)
- Right: Go to the next slide (Next)
- Up: Display all contents in the current window (Overview)
- Down: Copy/Paste the selected contents (Copy)

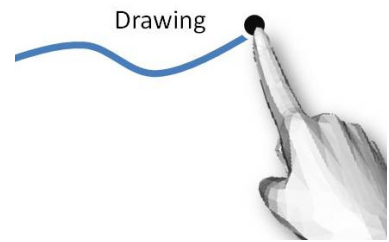

**Figure 2. Handwriting-based input (1-finger)**

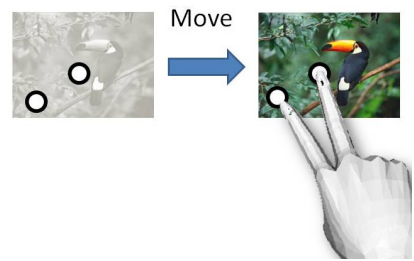

**Figure 3. Moving selected objects (2-fingers)**

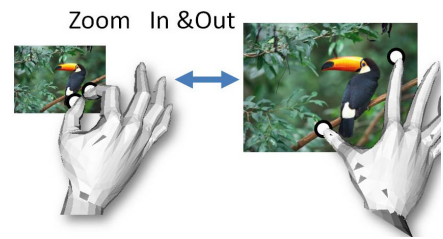

**Figure 4. Zoom in / Zoom out (2 fingers)**

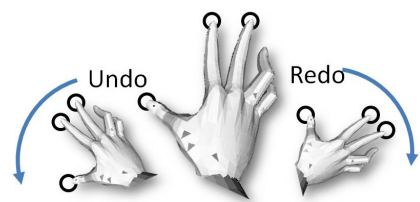

**Figure 5. Undo/Redo (3 fingers)**

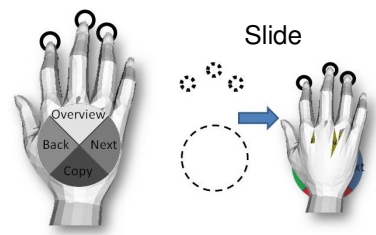

**Figure 6. The sPieMenu**

## EXPERIMENTS

In order to clarify the benefits of the multi-touch electronic whiteboard system described in the previous chapter, three sets of experiments were carried out. The first set of experiments looked at the number of menu items for the sPieMenu (abbreviated as SPM in this chapter only). The second set of experiments consisted of a usability comparison for the SPM and a traditional pie menu (abbreviated as PM in this chapter only). The third and final set of experiments compared the impact in terms of relative gaze movement for the proposed system versus the traditional electronic whiteboard system approach. These experiments are described below in order in detail.

**Number of Menu Items (Experiment 1)**

It is possible to support a large number of different functions using the proposed SPM approach by simply increasing the number of menu items in the pie menu. This however implies that the area and angle associated with each individual function will become smaller and thus require stricter, more accurate input movements in order to activate each function or menu item.

Here, an informal experiment was carried out in order to determine the number of different directions or items that could be reliably, accurately established with this input modality. This experiment evaluated the adequacy of the 4-item menu design described in the previous chapter, and involved 5 subjects.

*Method*

First the number of menu items for the SPM (the number of gesture angles) was set to 2, 4, 8, and 16 and each item was associated with an integer label. Figure 7 shows an example where the menu has been divided into 4 items. The test subjects were provided with a random integer label and asked to select it by inputting the appropriate direction. Each test subject was asked to perform 48 tests for each pie menu configuration and the average success rates for each configuration were then compared.

*Results*

The average success rates for the various configurations were as follows. For the 2 menu-item configuration the success rate was 99.6%, for the 4-item configuration the success rate was 98.3%, for the 8-item configuration 95.9%, and for the 16-item configuration 73.8%. Based on these results it was concluded that the 4-item or 8-item configurations were the most appropriate. Thus it was determined that the 4-item configuration described in the previous chapter was appropriate to utilize, and that it produced an acceptable error rate in the 1% region satisfying a "rigorous" decision-making standard.

In order to add additional functions or menu items to the pie menu some researchers have found that it is effective to add additional levels to the menu [5]. However adding multi-level functionality tends to also increase complexity. Expanding the set of available functions beyond a single list can also make the system more difficult for first-time users so this approach was not adopted in this paper.

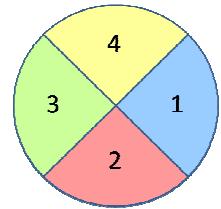

**Figure 7. Experimental menu configurations**

**PM and SPM Usability Comparison (Experiment 2)**

Electronic whiteboards are typically operated through interactions that are executed on one side of the screen. If the user attempts to interact with the whiteboard in the center of the screen he will end up with his back facing the audience and making eye-contact becomes difficult, while his body also partially obscures the presentation contents from the audience (Figure 8). Moreover, from the perspective of a new or first-time user, performing operations in the center of the screen makes it so that it is difficult to see the SPM menu using a natural posture, requiring the user to peek or strain to see the menu, and thus drastically decreasing usability. As a consequence of these two points, it is hard to justify supporting this sort of centralized interaction, and it is thus essential to focus on facilitating easy interactions when the user is standing at either side of the screen. Additionally in the event that the system is employed in a classroom setting, where an unspecified number of different users including teachers as well as students are expected to utilize the system, it is important to ensure that even novice users can operate the system without difficulty.

Thus the second experiment was designed in order to evaluate the impact on general usability for the PM and SPM systems, of the standing positions that novices as well as experienced users adopted with respect to the screen. The following two hypotheses were also proposed.

Hypothesis 1: From the perspective of experienced users, the SPM system should compare favorably with the PM system and not exhibit any drawbacks.

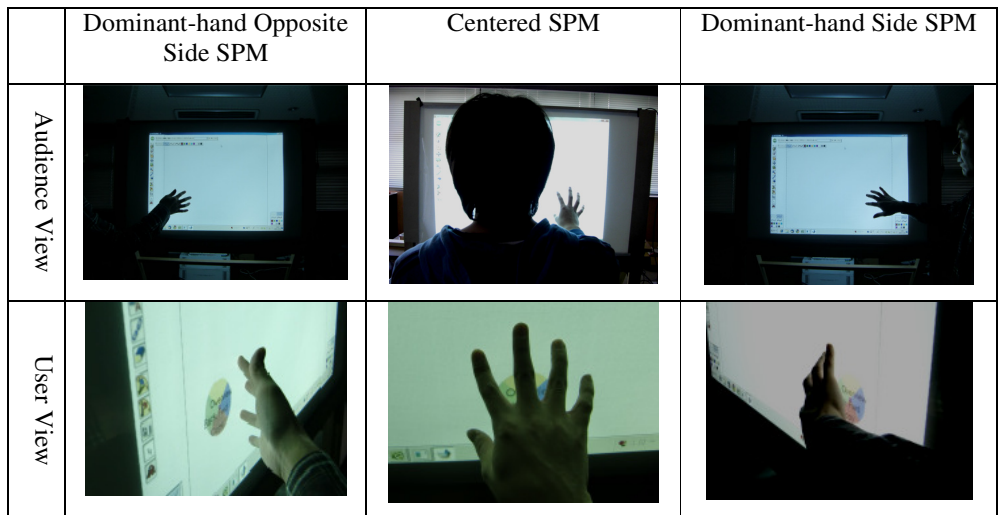

**Figure 8. Impact of standing position on user and audience views**

Hypothesis 2: From the novice user's perspective there will be some qualitative difference between the PM and SPM systems depending on the standing position adopted by the user.

*Method*

In this experiment, as with experiment 1, a reference number was associated with each menu item and the test subjects were asked to select a randomly chosen item. A program was constructed which presents a randomly selected menu item and then records the amount of time the user required to complete the action as well as whether or not they were able to complete it successfully. The pie menu utilized was the same as the one described in the previous experiment and displayed in Figure 7.

The experiment was set up as follows. Regarding the experience level designation of the user (novice/expert), users who had already memorized the positions and associated reference numbers for the menu items were considered expert users while those who had not yet memorized the reference numbers at all were considered novice users. Expert users were able to navigate the pie menu without looking at it, as they had already memorized the positions and associated reference numbers, while novice users on the other hand needed to look at the menu and descriptions each time they attempted a selection.

In order to accurately simulate this scenario, in the case of expert users the reference numbers were laid out as in Figure 7, in clockwise order, making it very simple for the test subjects to immediately establish an expert understanding of the menu items and associated reference numbers. In order to establish the novice user scenario, the reference numbers were randomly associated with the menu items in the pie menu for each test item, making it impossible for the test subjects to memorize the positions of the items.

Three potential standing positions were defined: Centered, Dominant-hand Dominant-side, Dominant-hand Opposite-side. In the case of the Dominant-hand Opposite-side position, this meant that a right-handed user would face the screen and stand on the left side (from the audience's perspective), while the Dominant-hand Dominant-side position meant that a right-handed user would face the screen and stand on the right side (from the audience's perspective) as shown in Figure 8. In the case of a left-handed user the standing positions would be reversed, but in these experiments all test subjects happened to be right-handed.

In the case of the Centered standing position, the user's body largely obscures the screen from the view of the audience, and thus this is not a practical position, however for the purposes of empirically comparison and consistency, this position was also included in the experiments.

Finally, regarding the menu display position (PM/SPM), the diameter of the pie menu was set so that it could be entirely covered by the palm of the test subject's hand. In the case of the PM configuration the menu was centered at the point where the test subject's middle finger touched the screen, while in the case of the SPM menu it was set to the center of the subject's palm.

The experiment involved 12 test subjects, each of whom were asked to test each of the 12 different configurations 20 times, for a total of 240 test cases per subject. In order to ensure that the subjects did not forget the menu arrangement, and thereby invalidate the expert user scenario, the first half of the experiments focused on the expert use case where the menu item configuration remained static, while the second half of the experiments simulated the novice condition. The standing position and menu display position were selected randomly throughout the experiments. Prior to carrying out the experiments all test subjects were thoroughly instructed in the use of the pie menu as well as the standing positions and experience levels, and were provided ample opportunity to practice using the system and remember the position and associated reference numbers for the menu items. For the purposes of analysis, as it was not difficult to imagine that the evaluations for the novice condition would take more time on average than those for the expert condition, the expertise levels were handled separately and the standing position as well as menu display position were each treated as independent variables while the task-completion time was treated as a dependent variable, resulting in a 3x2 analysis-of-variance table. The Bonferroni method was then employed to carry out a multiple comparison. With regard to the variance analysis, the times for successful completion as well as failure were considered together and a general comparison was carried out.

*Results*
Experimental results are described in Table 1 and Table 2.

**a. Overall successful completion rate**

The overall successful completion rate for this experiment was 98.5%.

**b. Expert condition**

The results of the variance analysis showed that the standing position had a significant main effect on the task completion time at ($F(2,1434) = 3.090, p<.05$) while the menu display position did not have a significant main effect. Additionally, there was no significant interaction that could be confirmed between the two conditions. Multiple comparison for the standing position showed that given a 5% significance level, there was not a significant difference.

**c. Novice condition**

The results of the variance analysis in this case showed that the main effect of the menu display position was significant, at ($F(1,1434)=35.379, p<.001$), while the main effect of the standing position was also significant at ($F(2,1434)=12.207, p<.001$), however there was also a significant interaction at ($F(2,1434)=32.515, p<.001$) implying that these main effects are qualified by the significant interaction.

Because the interaction was significant, tests of simple main effects were also conducted. The results of this analysis showed that in the Dominant-hand Dominant-side standing position, compared to the SPM condition, the PM condition was significantly faster($p<.001$). Also in the Dominant-hand Opposite-side standing position, SPM condition was significantly faster ($p<.01$), and in the case of the Centered standing position the PM condition was significantly faster ($p<.001$). Furthermore given the SPM condition, the Dominant-hand Opposite-side standing position was significantly faster than the Dominant-hand Dominant-side standing position with ($p<.001$). In addition, the Dominant-hand Dominant-side standing position was significantly faster than the Centered standing position with ($p<.01$), while similarly the Dominant-hand Opposite-side standing position was also significantly faster than the Centered standing position with ($p<.001$). Other conditions were not significant to within a standard 5% significance level.

*Discussion*
The expert users who have linked the 4-direction gesture and function menu in their memory were able to use the SPM menu without any problems despite the fact that the SPM menu would become invisible during use, which made it difficult to visibly confirm operations (Hypothesis 1). The results of analysis confirmed that the PM approach was not significantly faster and that from the perspective of the expert users the SPM approach was not lacking at all compared to the PM approach, which confirms Hypothesis 1 (see Table 1).

In the case of the novice users, because they need to be able to see the menu in order to successfully operate it, there is potential for the standing position to introduce a usability gap between the SPM and PM approaches (Hypothesis 2). The experimental results also supported Hypothesis 2, as described in Table 2. Looking more carefully at the analysis results showed that, for the Centered as well as the Dominant-hand Dominant-side standing conditions, the PM approach facilitated significantly faster operations than the SPM approach. However, in the case of the Dominant-hand Opposite-side standing position the SPM approach actually facilitated faster operations. This difference is most like due to the fact that in the case of the Dominant-hand Opposite-side standing position, the user is able to completely see the menu displayed under the palm of their hand, thus facilitating the operation by allowing the user to quickly confirm their desired actions even when he does not have the menu configuration memorized (Figure 8). In contrast, in the PM configuration the menu is usually partially obscured by the user's finger, thus preventing him from visually confirming some of the information, which may explain the lower average speed of these operations.

Moreover comments from the test subjects agreed with this analysis. In the case of the SPM approach the results for the Dominant-hand Dominant-side standing position were worse. Correspondingly multiple test subjects also stated that, "operations are more difficult from this side." In the case of a right-handed user this meant facing the screen, standing on the right side of the screen and then operating the system with thumb facing downward - essentially operating the system in an awkward 'backwards' manner. Accordingly, relatively large individual differences in the results among the subjects make it difficult to say whether the SPM approach is natural for all novice users. Namely, in the case of the Dominant-hand Dominant-side approach it was clear that, while some users were able to utilize the system very effectively, others were not.

Furthermore in the case of the novice users, the usability of the SPM condition changed for the various standing positions. This order was, from worst to best: Centered, Dominant-hand Opposite-side, Dominant-hand Dominant-side. Under the SPM condition the menu is shown under the palm of the user's hand so the novice, Centered, SPM combination ensured the worst possible combination of traits and the results reflected this. Related to this, many test subjects stated that this combination was the most difficult to see and operate. In the case of the Dominant-hand Dominant side standing position, the entire menu was not hidden from view, yet some portion of it often was. Finally, in the case of the Dominant-hand Opposite-side standing position, even though the menu was being displayed beneath the palm of the user's hand, it was easy to see the entire menu and associated functions, thus subjects stated that this was the easiest of the three standing positions to see.

The above discussion and analysis show that employing the SPM approach, which does not obscure the presentation materials, and combining this with a Dominant-hand Opposite-side standing position can provide a high degree of usability and effectiveness that is not dependent on the level of expertise of the user.

| **Expert User** | **PM Avg.(*SD*)** | **SPM Avg.(*SD*)** |
|---|---|---|
| Dom.-hand Dom.-side | 1.270(0.414) | 1.342(1.041) |
| Dom.-hand Opp.-side | 1.284(0.473) | 1.180(0.347) |
| Centered | 1.238(0.478) | 1.202(0.441) |

Table 1. Expert user condition: average task-completion time (seconds)

| **Novice** | **PM Avg.(*SD*)** | **SPM Avg.(*SD*)** |
|---|---|---|
| Dom.-hand Dom.-side | 1.578(0.458) | 1.799(0.550) |
| Dom.-hand Opp.-side | 1.668(0.555) | 1.534(0.341) |
| Centered | 1.563(0.516) | 1.984(0.737) |

Table 2. Novice user condition: average task-completion time (seconds)

**Audience Gaze Tracking (Experiment 3)**
In this experiment the single-touch toolbar which is currently in common use in many electronic whiteboard interfaces was compared with the newly proposed SPM multi-touch interface through a display task and the impact on audience gaze trajectory and overall gaze movement was measured to determine whether there was any difference between the two approaches. In this case the primary hypothesis was that the proposed interface would result in less unnecessary distraction and gaze movement for the audience compared to the traditional toolbar menu. That is to say, it should be possible to achieve effective "audience gaze localization" with the proposed method.

*Method*
Two types of simulated presentations were recorded using two interfaces. Each of them incorporated the following operations: picture object zoom (Figure 4); slide switching (Figure 6); and undo/redo (Figure 5). By showing the recorded videos, audience gaze tracking was performed on the experimental audience subjects.

The two interfaces consisted of the generic single-touch toolbar button interface from [4], which can be seen in Figure 9, and the proposed multi-touch SPM interface, which can be seen in Figure 10. In order to clearly determine the differences in audience gaze movement between the two interfaces, the materials intended as the focal point were always displayed and operated on near the center of the screen.

The recorded sample presentations were 90 seconds long for both interfaces, and the same functions were applied in the same order in both presentations.

The recorded video was interlaced, full hi-vision (1080i) quality, and was shown on a 37-inch hi-vision display. The test subjects were seated 140cm away from the screen and a gaze tracking system (EMR-9 Nac Inc.) was set up to record the experiments (Figure 11).

The toolbar button based presentation video was referred to as A, and the multi-touch SPM based presentation video as B. The test subjects were shown each video 2 times for a total of 4 viewings. In total 14 test subjects were shown the video presentations and the order of the video was switched between A, B, B, A and B, A, A, B for each subject so as to eliminate possible effect of the viewing order on the experimental results.

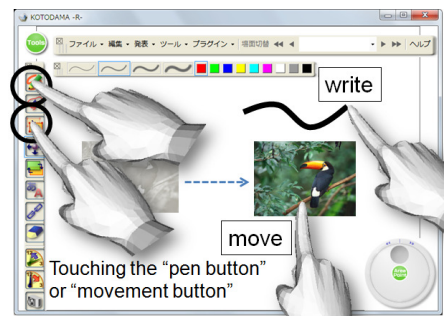

Figure 9. Toolbar button interface

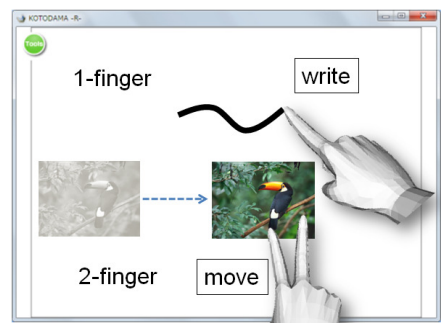

Figure 10. Multi-touch SPM interface

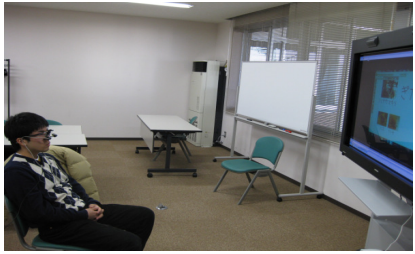

**Figure 11. Gaze tracking setup**

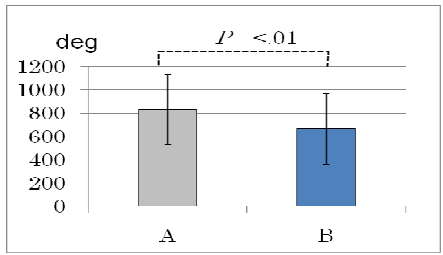

**Figure 12. Comparison of gaze movement for interfaces**

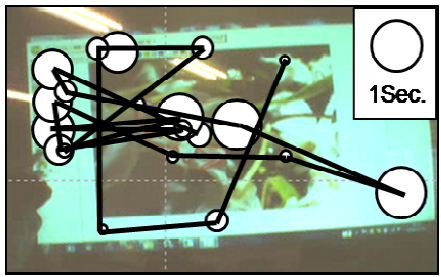

**Figure 13. Gaze tracking results for video A**

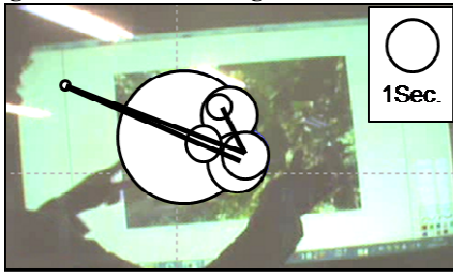

**Figure 14. Gaze tracking results for video B**

Gaze tracking results for each time-sample and each test subject (degree of gaze movement) were aggregated and computed and the experimental results are shown in Figure 12. Examining the relative eye movement associated with the two videos it is easy to see that there is a significant difference between the two results (t(13)= 3.7197, p<.01), and that B resulted in a significantly lower degree of gaze movement than condition A.

*Discussion*
The experimental results clearly showed that the proposed interface resulted in reduced gaze movement for the audience, which confirms the original hypothesis. Below, additional analysis regarding "where the audience looked" is also provided. Figure 13 and 14 describe the gaze movements of a single audience subject for the 'zoom' operation and track the subject's gaze movements as well as points of focus for video A and video B respectively. The straight lines describe gaze paths while the size of each focal point reflects the relative length of time the subject spent looking at that particular point.

Looking at the recordings for this test subject's gaze movement it can be seen that, in the case of A there is a large amount of movement, as well as a large number of different focus points (Figure 13). In the case of B, gaze movement is fairly limited, and characterized by long periods of focus on the center of the screen (Figure 14). Furthermore in A, there is particular focus on the left-hand side as well as the bottom right-hand side of the screen. Comparing the focal points of the subject's gaze with the overlay image (Figure 13) it is clear that the focal points on the left-hand side of the screen match up with the menu buttons, while those on the bottom right-hand side match up with the location of the zoom dial (Figure 9). On the other hand, in the case of B, where there is no visible tool button or dial (Figure 10), and function as well as menu related operations are carried out in the vicinity of the presentation contents, the test subject's gaze also focuses primarily on this contents area (Figure 14). Furthermore, these observations were consistent for all the test subjects.

The comments obtained from the test subjects such as, "Video A seemed to involve a lot more operational movement", "Video A was more busy and was tiring to watch", "Video B showed more coherent movement and seemed overall more smooth", "I was able to focus more clearly on the contents in the case of Video B", further support the above observations.

These results show that the gaze of the audience test subjects has a tendency to follow or track the movements or gestures of the presenter. This tendency may be the basis for the higher rate of visual recognition that has been observed for direct touch interfaces versus traditional mouse or pointer based interfaces [2], so it should not present a problem in this context. Tracking gazes include two different types, "Meaningful tracking gazes" where the audience's gaze is following movements intended to explain or point out the actual presentation materials, and "meaningless tracking gazes" where the audience's gaze is tracking movement that serves only to select items from a menu or select or deselect certain functions. Here it is important to try and minimize the occurrence of the latter "meaningless tracking gazes". At the very least it is clear that audience focus on peripheral menus, etc. that are unrelated to the actual teaching or presentation materials is not the purpose of any classroom lesson, and the proposed interface was clearly shown to reduce the prevalence of "meaningless tracking gazes".

Nevertheless even when the presenter's hand movements are focused in the vicinity of the presentation content, this does not necessarily guarantee that the audience is actually focusing on that content. Sometimes the audience may simply be distracted by the presenter, and lose focus on the presentation content itself. Regardless of the actual intentions of the presenter, it is often also the case that the movements of the presenter end up being combined with the presentation materials and received by the audience as a hybrid multimedia content experience. Thus the issue of the importance or lack thereof of the presenter's hand motions is not a trivial one. At the least, a system that supports an interaction paradigm where the user's operations can be wholly carried out in the immediate vicinity of the actual presentation content is unlikely to prove detrimental to such a combined multimedia presentation approach.

Related to this, it is important to evaluate the reactions of audiences to the various movements and operations to which they are exposed. For example, if a PM approach to a menu display interface is employed, each operation obscures the presentation contents and the audience's attention is captured not by the presentation materials or the presenter but by the menu interaction itself. The SPM approach however, being designed to ensure that presentation contents are not obscured, and that the menu itself is generally hidden from the audience's view is more effective at eliminating these barriers compared with the PM approach. Nevertheless we cannot say that the menu will not occasionally be visible between the presenter's fingers, nor that unnatural gesture input will not sometimes distract or steal the attention of the audience.

In order to perform a more detailed verification regarding the effect of such interactions on audience attention it may be useful to perform a larger 3-way cross-comparison of (1) an automated video presentation where no menu is shown, (2) SPM, and (3) PM, looking again at relative gaze movement and relative levels of comprehension. Moreover from the perspective of an actual implementation it is important to carefully investigate the impact of the way functions are associated with menu items or gestures. It is also important to separate actions or items that should be deliberately shown to the audience in order to improve understandability, from those that should perhaps not be disclosed. We leave these ideas for future work.

**CONCLUSION**

This research developed a new electronic whiteboard system that employs a multi-touch gesture interface and sPieMenu, and made a new effort to achieve audience gaze localization. Several evaluation experiments were carried out and the following results were obtained. (1) It was shown that the proposed approach reduced the degree of audience gaze movement. (2) In the case of a "Dominant-hand Dominant-side" standing position (for a right-handed subject this means standing on the left side of the screen when facing it) this approach facilitated as fast or faster interactions.

In future, it will be necessary to research how and to what degree interface driven changes in gaze movement impact audience comprehension and focus. Additionally, utilizing the space where the sPieMenu was displayed as a private space for the presenter and investigating how this space might be designed or utilized to further support the presenter may represent an interesting field for future research.

**REFERENCES**

1. Callahan, J., Hopkins, D., Weiser, M. and Shneiderman, B. "An Empirical comparison of Pie Versus Linear Menus," *In Proc. of ACM SIGCHI'88*, pp.95-100, 1988.
2. Elrod, S., Brouce, R., Gold, R., Goldberg, D., Halasz, F., Janssen, W., Lee, D., McCall, K., Pedersen, E., Pier, K., Tang, J. and Welch, B. "Live board: A Large Interactive Display Supporting Group Meetings, Presentations and Remote Collaboration," *In Proc. of ACM SIGCHI'92*, pp.599-607, 1992.
3. Han, J. Y. "Low-Cost Multi-Touch Sensing through Frustrated Total Internal Reflection," *In Proc. of ACM UIST'05*, pp.115-118, 2005.
4. Kurihara, K., Igarashi, T. and Ito, K. "A Pen-based Presentation Tool with a Unified Interface for Preparing and Presenting and Its Application to Education Field," *Computer Software*, Vol.23, No.4, pp.14-25, 2006.
5. Kobayashi, M. and Igarashi, T. "Considering the direction of cursor movement for efficient traversal of cascading menus," *In Proc. of* UIST'03, pp.91-94, 2003.
6. Norman, D. A. "The Psychology of Everyday Things," Basic Books, 1988.
7. Oguni, T. and Nakagawa, M. "Prototyping of various applications for an interactive electronic white board," *JSPS Technical Report 96-HI-62*, pp.9-16, 1996.
8. Brandl, P., Leitner, J., Seifried, T., Haller, M., Doray, B. and To, P. "Occlusion-Aware Menu Design for Digital Tabletops," *In Proc of ACM SIGCHI'09*, pp.3223-3228, 2009.
9. Vogel, D., Cudmore, M., Casiez, G., Balakrishnan, R. and Keliher, L. "Hand Occlusion with Tablet-sized Direct Pen Input," *In Proc of ACM SIGCHI'09*, pp.557-566, 2009.
10. Gonzalo Ramos, Matthew Boulos, and Ravin Balakrishnan, "Pressure Widgets," *In Proc. of ACM SIGCHI'04*, pp.487-494, 2004.
11. Tovi Grossman, Ken Hinckley, Patrick Baudisch, Maneesh Agrawala, and Ravin Balakrishnan, "Hover Widgets: Using the Tracking State to Extend the Capabilities of Pen-operated Devices," *In Proc. of ACM SIGCHI'06*, pp.861-870, 2006.
12. LAO, S., HENG, X., ZHANG, G., LING, Y. and WANG, P. (2009) A Gestural Interaction Design Model for Multi-touch Displays. *British Computer Society Conference on Human-Computer Interaction* : 440-446
13. Kurtenbach, G. and Buxton, W. "User Learning and Performance with Marking Menus," *In Proceedings of the SIGCHI conference on Human factors in computing systems*, pp. 258-264, ACM Press, 1994.